# MUCM-Net: A Mamba Powered UCM-Net for Skin Lesion Segmentation

Chunyu Yuan[1], Dongfang Zhao[2], Sos S. Agaian[1,3]

1. The Graduate Center, City University of New York
2. University of Washington
3. College of Staten Island, City University of New York

**Abstract** – Skin lesion segmentation is key for early skin cancer detection. Challenges in automatic segmentation from dermoscopic images include variations in color, texture, and artifacts of indistinct lesion boundaries. Deep learning methods like CNNs and U-Net have shown promise in addressing these issues. To further aid early diagnosis, especially on mobile devices with limited computing power, we present MUCM-Net. This efficient model combines Mamba State-Space Models with our UCM-Net architecture for improved feature learning and segmentation. MUCM-Net's Mamba-UCM Layer is optimized for mobile deployment, offering high accuracy with low computational needs. Tested on ISIC datasets, it outperforms other methods in accuracy and computational efficiency, making it a scalable tool for early detection in settings with limited resources. Our MUCM-Net source code is available for research and collaboration, supporting advances in mobile health diagnostics and the fight against skin cancer. In order to facilitate accessibility and further research in the field, the MUCM-Net source code is https://github.com/chunyuyuan/MUCM-Net

## 1 Introduction

Skin cancer remains a major health concern worldwide, ranking among the top diagnosed cancers. It is generally categorized into two primary types: melanoma and non-melanoma. Melanoma, although less common, comprising just 1% of skin cancer cases, disproportionately causes the majority of skin cancer-related deaths due to its aggressive nature. In the United States, melanoma was responsible for an estimated 7,800 deaths in 2022, with new cases projected to reach 98,000 in 2023 [1]. The lifetime risk of developing skin cancer for Americans is significant, with current data indicating that one in five will be affected, highlighting the urgent need for effective diagnostic and treatment strategies. The financial burden is also substantial, with skin cancer treatment costs in the U.S. estimated at over $ 8.1 billion in the United States alone [2]. Skin cancer, particularly malignant melanoma, is known for its swift progression and high mortality rate, making early and accurate diagnosis crucial for enhancing patient outcomes [3]. Dermatoscopy and dermoscopy are pivotal in the clinical assessment of skin lesions, aiding dermatologists in identifying malignant features [4]. However, manual interpretation can be time-consuming and error-prone, dependent on the clinician's expertise. Recent advancements have introduced machine learning-driven techniques into clinical practice to improve diagnosis accuracy and efficiency. These techniques are particularly beneficial in computationally constrained environments like mobile health applications [5, 6].

Manual interpretation can be time-consuming, error-prone, and heavily dependent on the clinician's expertise. Additionally, specific medical samples pose significant challenges [44]:

unclear boundaries where lesions blend into surrounding skin; illumination variations altering lesion appearance; artifacts like hair and bubbles obscuring lesion boundaries; variability in lesion size and shape; differences in imaging conditions and resolutions; age-related skin changes affecting texture; complex backgrounds hindering segmentation; and differences in skin color due to race and climate. Figure 1 shows representative samples of complex skin lesions.

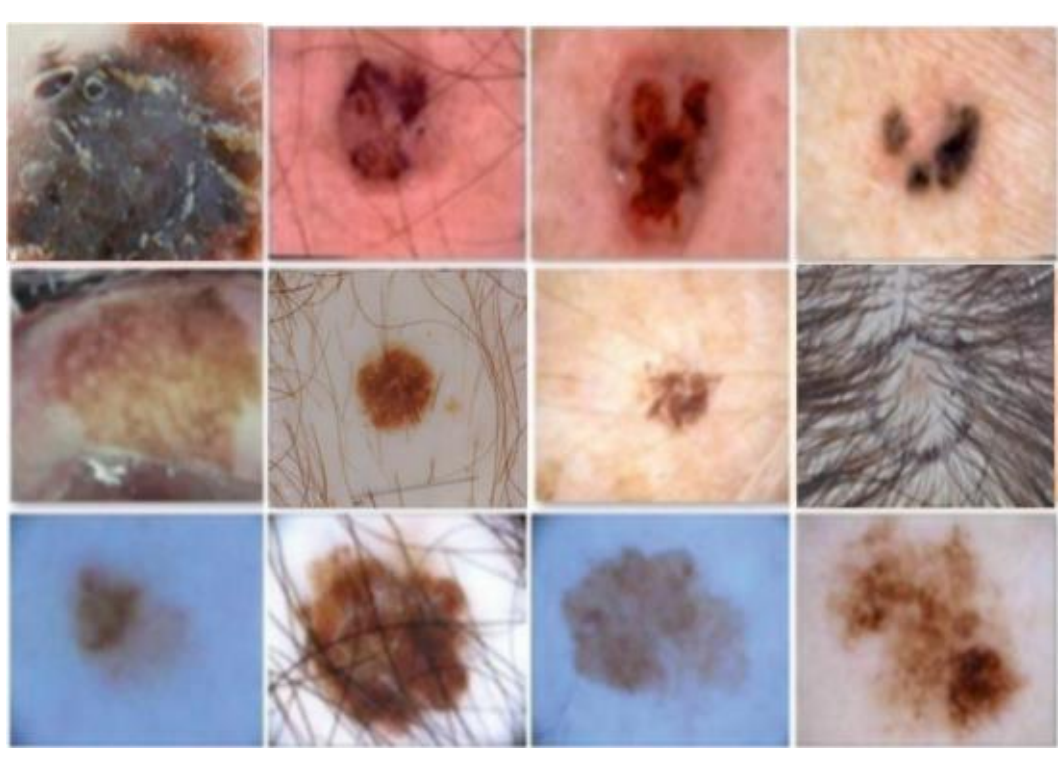

Figure 1: Complex skin lesion samples

To enhance the precision and efficiency of skin cancer diagnosis, recent advancements have increasingly incorporated computer-aided tools and artificial intelligence (AI) into clinical practice [7,8]. A critical technique in this domain is skin cancer segmentation, which precisely identifies the margins of skin lesions in medical images. This segmentation is vital for accurately assessing lesion characteristics, monitoring their progression, and guiding treatment decisions. With rapid advancements in AI techniques and the widespread adoption of smart devices, such as point-of-care ultrasound (POCUS) devices or smartphones [9, 10, 11], AI-driven approaches for skin cancer detection have become popular.

Patients now enjoy enhanced access to medical information, remote monitoring, and tailored care, which has improved their overall satisfaction with healthcare services. Despite these positive changes, certain obstacles remain, particularly in medical diagnostics. A notable issue is the precise and efficient segmentation of skin lesions, which is critical for diagnosis but challenging to implement on devices with limited computational resources. Most AI-driven medical applications rely on deep learning techniques described in detail by [12]. These methods typically require significant computational power and extensive learning parameters to deliver accurate predictions, posing a challenge for integration into devices with constrained hardware capabilities [13, 14].

State-space models (SSMs) have recently been recognized for their linear complexity concerning input size and memory usage, establishing them as fundamental components for lightweight model architectures [15]. SSMs are particularly effective at capturing long-range dependencies, offering a critical solution to the convolution challenge of processing information across extensive distances. With the advantage of SSMs, Mamba [16] has been proven to handle textual data with fewer parameters than Transformers. Similarly, the advent of Vision Mamba [17] has advanced the application of SSMs in image processing, demonstrating a significant memory reduction, all without relying on traditional attention mechanisms. This pioneering research bolsters confidence in Mamba's potential as a critical lightweight model component in future technological advancements.

In this study, we extend our previous method UCM-Net [18], and introduce MUCM-Net, a lightweight, robust and mamba-powered approach for skin lesion segmentation. MUCM-Net leverages a new novel hybrid module that combines Convolutional Neural Networks (CNN), Multi-Layer Perceptions (MLP) and Mamba to enhance feature learning. Utilizing new proposed group loss functions, our method surpasses existing mamba-based techniques in skin lesion segmentation.

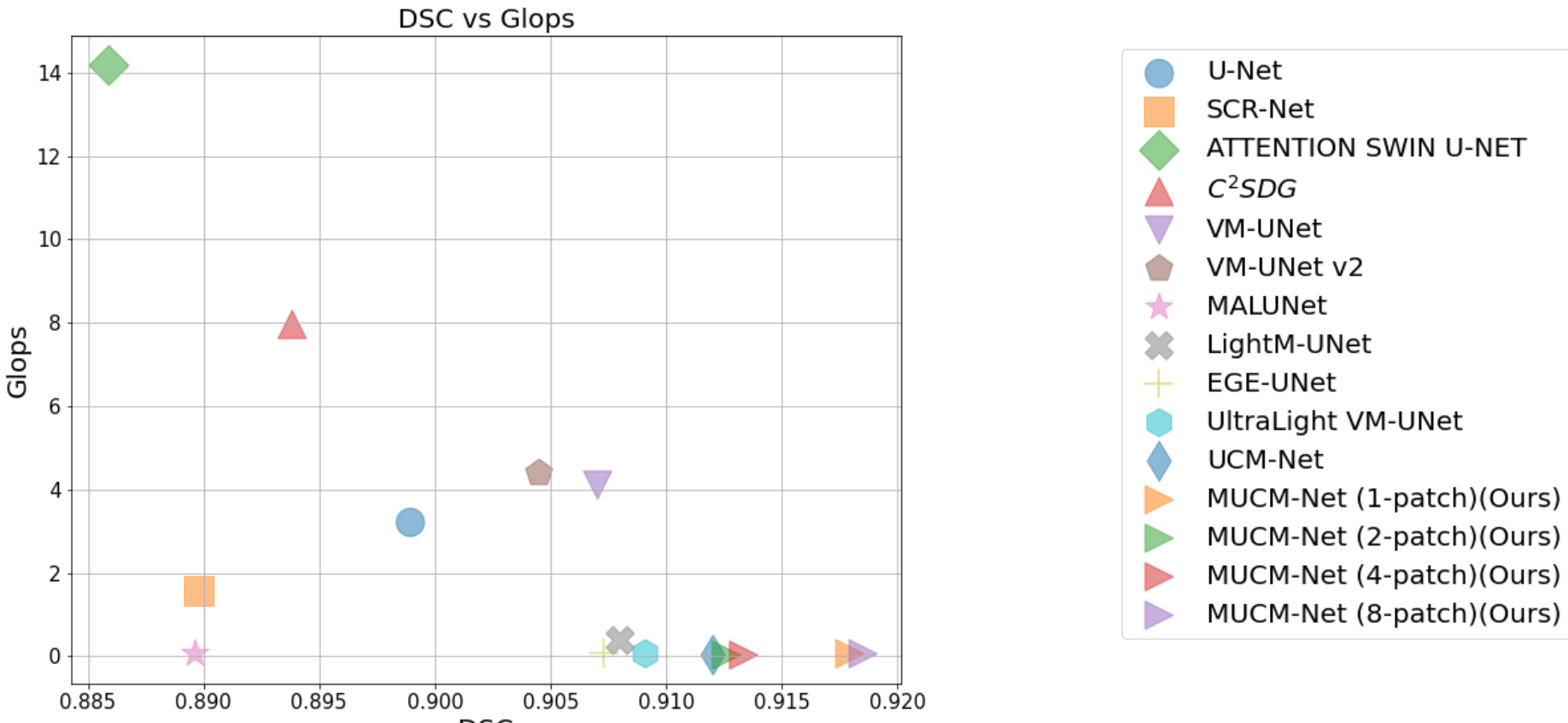

Figure 2: This figure shows the visualization of comparative experimental results on the ISIC2017 dataset. The X-axis represents DSC (higher is better), while Y-axis represents Glops (lower is better)

Key contributions of MUCM-Net include:

- **Hybrid Feature Learning**: The MUCM-Net Block integrates CNN, MLP, and Mamba elements, enhancing the learning of complex and distinct lesion features.
- **Computational Efficiency**: MUCM-Net's design, based on Mamba-UCM Blocks and UCM-Net, prioritizes accuracy and efficiency. It achieves high prediction performance with low computational demands (approx. 0.055-0.064 GFLOPs), making it suitable for various deployment scenarios.
- **Enhanced Loss Function**: A novel loss function integrates output and internal stage losses, ensuring efficient learning during the model's training process.
- **Superior Results**: MUCM-Net achieves exceptional results on the ISIC 2017 and 2018 datasets, outperforming previous Mamba-based methods on metrics like Dice similarity, sensitivity, specificity, and accuracy.

## 2 Related Work

### 2.1 TinyML for Healthcare

Biomedical imaging segmentation involves precisely delineating anatomical structures and pathological regions from medical images, is critical for accurate diagnostics. Recent strides in artificial intelligence (AI) have significantly advanced segmentation techniques, greatly enhancing their accuracy and efficiency.

An emerging frontier in this domain is the integration of TinyML into healthcare, particularly for tasks such as lesion segmentation, which offers promising research avenues and practical applications. TinyML refers to implementing machine learning models on low-power, compact hardware. This technology can potentially revolutionize healthcare by bringing advanced analytical capabilities directly to the point of care. It enables real-time, on-device processing, making sophisticated medical image analysis accessible even in environments with limited traditional computing resources or mobile healthcare settings. For example, leveraging TinyML for lesion segmentation could provide immediate diagnostic insights during patient examinations or in remote areas, dramatically reducing the reliance on extensive infrastructure typically required for detailed analyses. The integration of TinyML into medical devices is poised to improve diagnostic processes, enhance patient outcomes, and expand the availability of advanced medical technologies to underserved areas. To maximize the efficiency and feasibility of deploying TinyML in such critical applications, researchers are investigating advanced techniques like hyper-structure optimization [19] and employing quantitative methods such as binary neural networks [20].

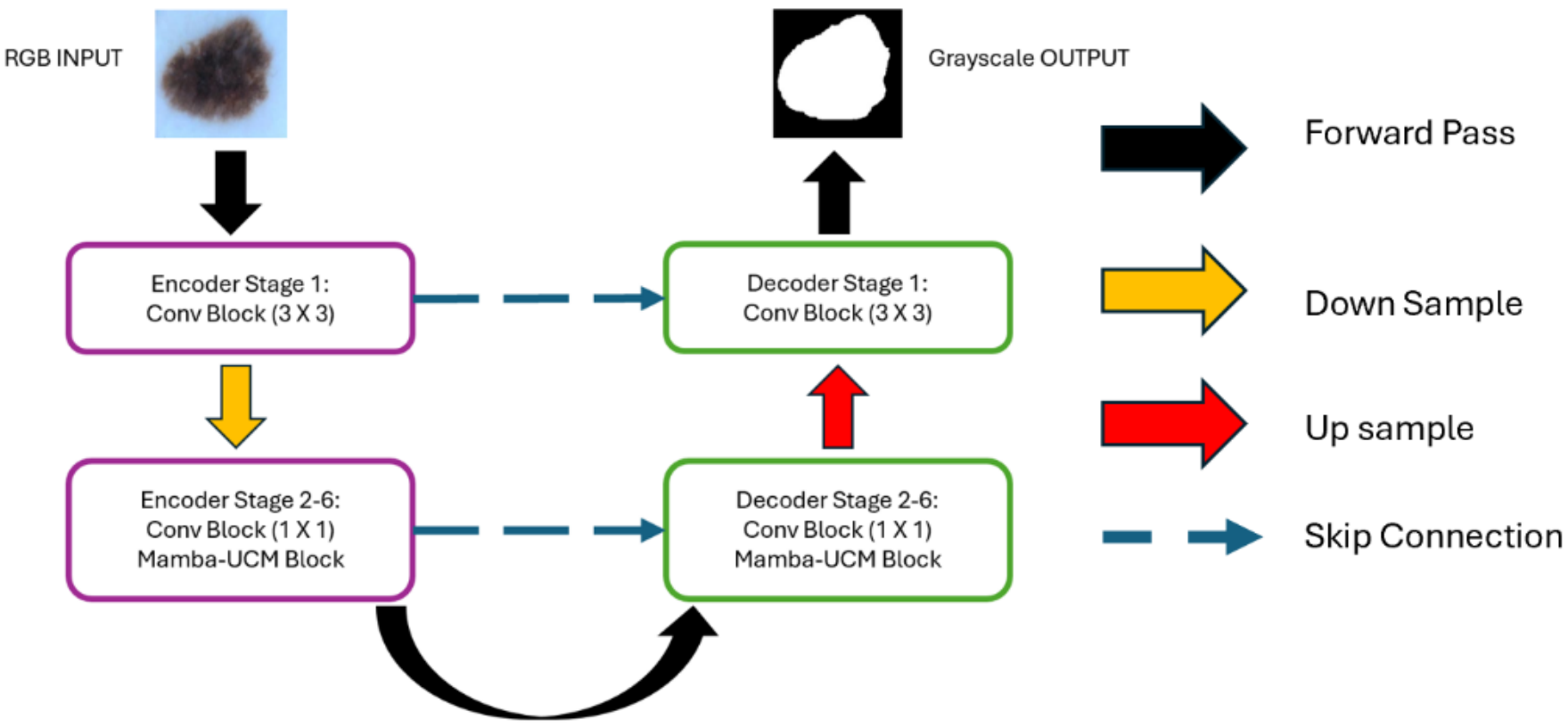

Figure 3: MUCM-Net Structure

Hyper-structure optimization focuses on reducing the model's parameter count without sacrificing performance, ensuring the models remain both practical and lightweight for use on miniature devices. Moreover, implementing binary neural networks helps streamline computations, further enhancing the practicality of TinyML applications in resource-constrained settings. As we delve into optimizing and applying models like MUCM-Net in healthcare. This research not only underscores the transformative possibilities of TinyML but also guides future explorations in deploying compact, efficient AI solutions in medical settings.

### 2.2 Supervised Methods of Segmentation

As AI technology continues to advance, the approaches for medical image segmentation have evolved significantly. Initially, the field heavily relied on convolutional neural networks (CNNs) such as U-Net and its attention-enhanced variant, Att-UNet [], which incorporates attention mechanisms to further refine the segmentation accuracy by focusing on relevant features within the images. The development of hybrid architectures marks a further evolution in segmentation techniques. There are some hybrid-based UNets for medical image segmentation: (1) Transformers-related: such as TransUNet [21], TransFuse [22] and SANet [23]; (2) multilayer perceptron (MLP)-related: such as ConvNeXts [24], ConvNeXts [24], UNeXt [25], MALUNet [26] and its extended version EGE-UNet [27]. Recently, as Vision Mamba [17]'s image processing ability with fewer parameters and lower computations, Mamba-based hybrid structure UNets are becoming popular such as VM-UNet [28], VM-UNet V2 [29], LightM-UNet [30] and UltraLight VM-UNet [31].

In this paper, we extend our previous work UCM-Net [18] to propose a new hybrid work MUCM-Net which engages the Mamba's features learning ability and maintain fewer parameters and lower computations.

## 3 MUCM-Net

3.1 Network structure Design

Figure 3 provides a comprehensive view of the structural framework of MUCM-Net, an advanced architecture that showcases a distinctive U-Shape design. Our design is developed from UCM-Net. MUCM-Net includes a down-sampling encoder and an up-sampling decoder, resulting in a high-powered network for skin lesion segmentation. The entirety of the network encompasses six stages of encoder-decoder units, each equipped with channel capacities of {8, 16, 24, 32, 48, 64}. Within each stage, the first encoder-decoder stage is a convolutional block, which facilitates the extraction and acquisition of essential features. The rest of the stages are alongside our novel UCMNet blocks.

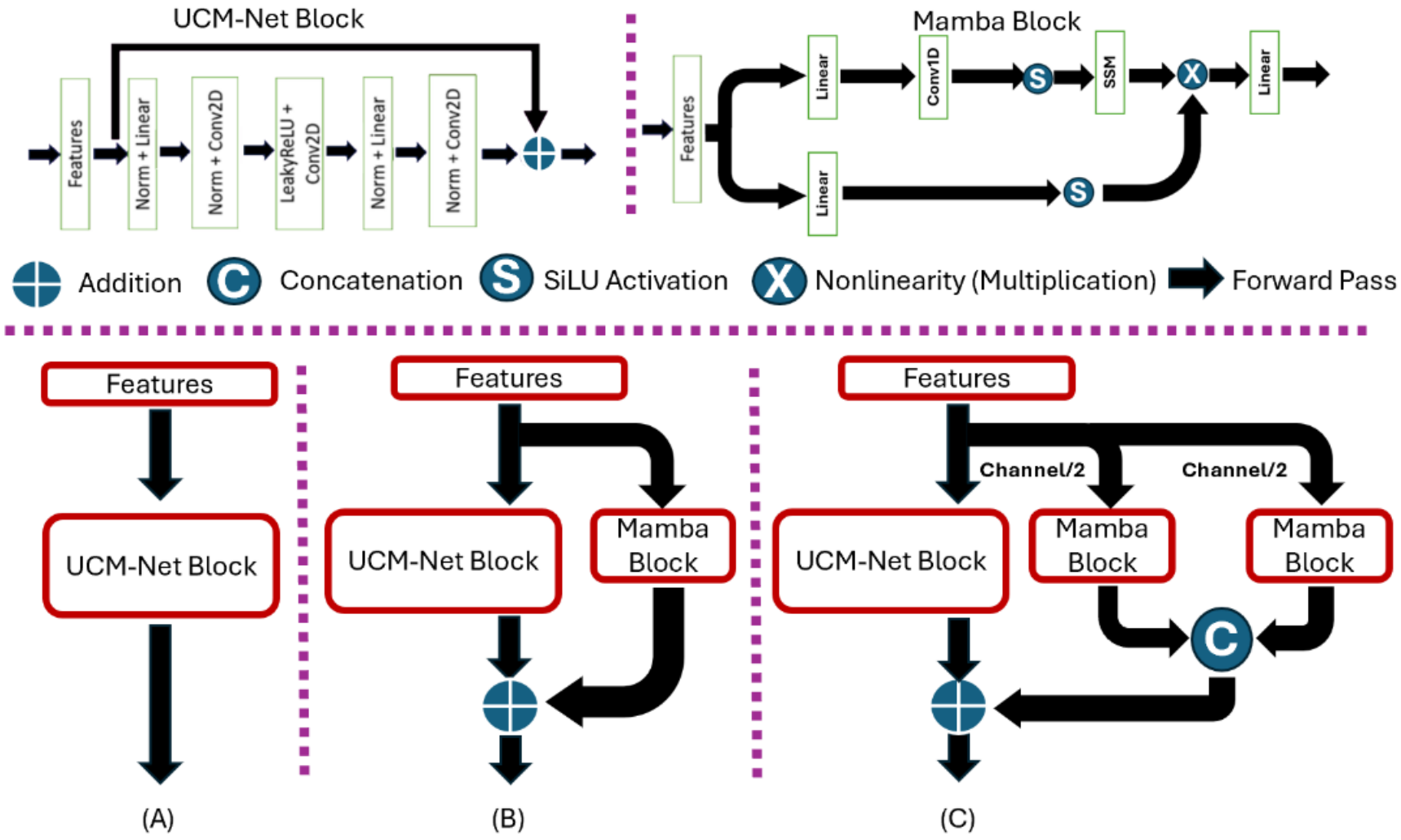

Figure 4: MUCM-Net Structure: (A) UCM-Net Pipeline, (B) MUCM-Net(1-patch) Pipeline, (C) MUCM-Net(2-patch) Pipeline

### 3.2 **Convolution Block**

The first encoder-decoder stage uses a standard convolution layer with a 3x3 filter in our design. Convolution Block utilizes a kernel size of 3×3, which is commonly employed to capture spatial relationships within the input features. This size is particularly advantageous in the network's initial layers, where preserving the spatial integrity of feature maps is essential for decoding complex input patterns. In the 2nd-6th stage, we use a 1x1 filter convolution layer to service the later Mamba-UCM block. To drastically reduce the number of learnable parameters and computational load.

### 3.2 **Mamba-UCM Block**

The 2nd-6th stages mainly use the Mamba-UCM block for feature learning. The Mamba-UCM Block showcases an advanced strategy that merges UCM-Net Block, which contains Convolutional Neural Networks (CNNs) with Multilayer Perceptions (MLPs) and Mamba Block, to enhance feature learning. This hybrid model leverages the spatial feature extraction

```
Algorithm 1 PyTorch-style pseudocode for Mamba-UCM Block(2-patch)
# Input:  X,the feature map with shape [Batch(B), Channel(C), Height(H), Width(W)]
# Output:  Out,the feature map with shape [B, Height*Width(N),C]
# Operator:  Conv, 2D Convolution  LN, LayerNorm  BN, BatchNorm,  Linear, Linear
Transformation  Mamba1, Mamba Layer,  Mamba2, Mamba Layer,  Leaky, Leaky RelU
# Mamaba-UCM Block Processing Pipeline
# *********************************
# Input X
# *********************************
B, C, H, W = X.shape()
# Transform Feature from [B,C,H,W] to [B,H*W,C]
X = X.flatten(2).transpose(1,2)
# Copy feature for later residual addition
X1 = copy(X)
# Copy feature for later Mamba
X2 = copy(X)
# *********************************
# UCM-Net block
# *********************************
X = Linear(LN(X))
B, N, C = X.shape()
# Transform Feature from [B,H*W,C] to [B,C,H,W]
X = X.transpose(1,2).view(B,C,H,W)
X = Conv(LN(X))
X = Conv(Leaky(X))
# Transform Feature from [B,C,H,W] to [B,H*W,C]
X = X.flatten(2).trnaspose(1,2)
X = Linear(BN(X))
# Transform Feature from [B,H*W,C] to [B,C,H,W]
X = X.transpose(1,2).view(B,C,H,W)
X = Conv(LN(X))
# Transform Feature from [B,C,H,W] to [B,H*W,C]
X = X.flatten(2).transpose(1,2)
# *********************************
# Mamba patch
# *********************************
# Based on channel to split into two patches
halves = torch.split(X2, X2.shape[2] // 2, dim=2)
# Mamba Layer process for each patch
x2-mamba-output1 = Mamba1(halves[0])
x2-mamba-output2 = Mamba1(halves[0])
# Concatenate two patch output
X2 = torch.cat((x2-mamba-output1, x2-mamba-output2), dim=2)
# *********************************
# Output Out
# *********************************
# Output with residual addition
Out = X + X1 + X2
```

Figure 5: Mamba-UCM Block Pseudocode

strengths of CNNs and the pattern recognition capabilities of MLPs and SSM. The process begins by reshaping the initial input feature map to meet the distinct requirements of CNNs,

MLPs, and Mamba. This adaptation involves converting a four-dimensional tensor suitable for CNN processing into a three-dimensional tensor appropriate for MLP and Mamba operations. Inspired by UltraLight VM-UNet [17] and Vision Mamba [32], we proposed four versions of MUCM-Net with different patch processing. Figure 4 presents the visible differences between the UCM-Net block and two visions of Mamba-UCM blocks. The PyTorch-Style pseudocode in Figure 5 presents our defined sequence of operations, which is how we combine the UCM block and the Mamba block for feature learning.

### 3.3 **Loss Function**

In our solution, we designed a new group loss function similar to those used in TransFuse [22], EGE-UNet [27], and our previous work, UCM-Net [18]. However, different from theirs, our proposed base loss function is calculated from binary cross-entropy (BCE) (1) and Dice Loss (2) components and Squared Dice Loss (3) components to calculate the loss from the scaled layer masks in different stages compared with the ground truth masks. Equations (5) and (6) present the stage loss in different layers and the output loss in the output layer, which is calculated using binary cross-entropy (BCE) and Dice loss (Dice) components, respectively.

$$BCE = -\frac{1}{N}\sum_{i=1}^{N}[y_i\, log(p_i) \; + (1-y_i)\, log(1-p_i)] \quad (1)$$

where *N* is the total number of pixels (for image segmentation) or elements (for other tasks), $\mathrm{y_i}$ is the ground truth value, and $\mathrm{p_i}$ is the predicted probability for the i-th element.

$$\text{Dice Loss} = 1 - \frac{2 \times \sum_{i=1}^{N}(p_i \cdot y_i) + \text{smooth}}{\sum_{i=1}^{N} p_i + \sum_{i=1}^{N} y_i + \text{smooth}} \quad (2)$$

where smooth is a small constant added to improve numerical stability.

$$\text{Squared Dice Loss} = 1 - \frac{2\times\left(\sum_{i=1}^{N}(p_i \cdot y_i)\right)^2 + \text{smooth}}{\left(\sum_{i=1}^{N} p_i\right)^2 + \left(\sum_{i=1}^{N} y_i\right)^2 + \text{smooth}} \quad (3)$$

which represents an enhancement over the standard Dice loss by emphasizing the squared terms of intersections and union.

$$\text{Base_loss} = \text{BCE} + \text{Dice Loss} + \text{Squared Dice Loss} \quad (4)$$

Equations (1), (2) and (3) define the base loss function (4) for our proposed model, incorporating the Dice loss, and squared-Dice loss components. $\lambda_i$ is the weight for different stages. In this paper, we set $\lambda_i$ to 0.1, 0.2, 0.3, 0.4, and 0.5 based on the i-th stage, as illustrated in Figure 4. Equation 7 is our proposed group loss function that calculates the loss from the scaled layer masks in different stages with ground truth masks. Equations 5,6 present the stage loss in different stage layer and output loss in the output layer.

$$Loss_{Stage} = \text{Base_loss}(StagePred, Target) \quad (5)$$

$$\text{Loss}_{\text{Output}} = \text{Base_loss(OutputPred,Target)} \quad (6)$$

$$Group_Loss = Loss_{Output} + \sum_{i=1}^{5} \lambda_i \times Loss_{Stage_i} \quad (7)$$

## 4 RESULTS AND DISCUSSION

### 4.1 DATASET

To evaluate the efficiency and performance of our proposed model with other published models, we pick the two public skin segmentation datasets from the International Skin Imaging Collaboration, namely ISIC2017 [32, 34] and ISIC2018 [35, 36]. The ISIC2017 dataset comprises 2000 dermoscopy images, and ISIC2018 includes 2594 images. The ISIC2017 dataset was randomly divided into 1250 for training, 150 for validation, and 600 for testing. The ISIC2018 dataset was randomly divided into 1815 for training, 259 for validation, and 520 for testing.

### 4.2 EVALUATION SETTING

Our MUCM-Net is implemented with the PyTorch [36] framework. All experiments are conducted on the instance node at Lambda [37] that has a single NVIDIA RTX A6000 GPU (24 GB), 14vCPUs, 46 GiB RAM, and 512 GiB SSD. The images are normalized and resized to 256×256. Simple data augmentations are applied, including horizontal flipping, vertical flipping, and random rotation. We select AdamW [38] for the optimizer, initialized with a learning rate of 0.001 and a weight decay of 0.01. The CosineAnnealingLR [39] is Utilized as the scheduler with a maximum number of iterations of 50 and a minimum learning rate of 1e-5. A total of 200 epochs are trained with a training batch size of 8 and a testing batch size of 1.

### 4.3 EVALUATION METRICS

The model's performance is evaluated using the Dice Similarity Coefficient (DSC), sensitivity (SE), specificity (SP), and accuracy (ACC). Furthermore, the model's memory consumption is assessed based on the number of parameters and Gigaflops (GFLOPs). DSC measures the degree of similarity between the ground truth and the predicted segmentation map. SE is used to measure the percentage of true positives in relation to the sum of true positives and false negatives. SP measures the percentage of true negatives in relation to the sum of true negatives and false positives. ACC measures the overall percentage of correct classifications. The formulas used are as follows:

$$\text{DSC} = \frac{2 \times TP}{2 \times TP + FP + FN} \quad (8)$$

$$\text{ACC} = \frac{TP + TN}{TP + TN + FP + FN} \quad (9)$$

$$\text{SE} = \frac{TP}{TP + FN} \quad (10)$$

$$\text{SP} = \frac{TN}{TN + FP} \quad (11)$$

where TP denotes true positive, TN denotes true negative, FP denotes false positive, and FN denotes false negative.

In our benchmark experiments, we evaluate our method's performance and compare the results among other published efficient models'. To ensure a fair comparison, we perform three sets of experiments for each method and subsequently present the mean and std of the prediction outcomes across each dataset.

| Dataset | Models | Year | DSC↑ | SE↑ | SP↑ | ACC↑ |
|---|---|---|---|---|---|---|
| isic2017 | U-Net | 2015 | 0.8989 | 0.8793 | 0.9812 | 0.9613 |
| | SCR-Net | 2021 | 0.8898 | 0.8497 | 0.9853 | 0.9588 |
| | ATTENTION SWIN U-NET | 2022 | 0.8859 | 0.8492 | 0.9847 | 0.9591 |
| | $C^2SDG$ | 2021 | 0.8938 | 0.8859 | 0.9765 | 0.9588 |
| | VM-UNet | 2024 | 0.9070 | 0.8837 | 0.9842 | 0.9645 |
| | VM-UNet v2 | 2024 | 0.9045 | 0.8768 | 0.9849 | 0.9637 |
| | MALUNet | 2022 | 0.8896 | 0.8824 | 0.9762 | 0.9583 |
| | LightM-UNet | 2024 | 0.9080 | 0.8839 | 0.9846 | 0.9649 |
| | EGE-UNet | 2023 | 0.9073 | 0.8931 | 0.9816 | 0.9642 |
| | UltraLight VM-UNet | 2024 | 0.9091 | 0.9053 | 0.9790 | 0.9646 |
| | UCM-Net (Baseline) | 2024 | 0.9120 | 0.8824 | 0.9877 | 0.9678 |
| | MUCM-Net (1-patch) | 2024 | 0.9179 | 0.9089 | 0.9833 | 0.9692 |
| | MUCM-Net (2-patch) | 2024 | 0.9126 | 0.9008 | 0.9829 | 0.9673 |
| | MUCM-Net (4-patch) | 2024 | 0.9133 | 0.8871 | 0.9870 | 0.9681 |
| | MUCM-Net (8-patch) | 2024 | 0.9185 | 0.9014 | 0.9857 | 0.9697 |

Table 1: Comparative prediction results on the ISIC2017 datase

| Dataset | Models | Year | DSC↑ | SE↑ | SP↑ | ACC↑ |
|---|---|---|---|---|---|---|
| isic2018 | U-Net | 2015 | 0.8851 | 0.8735 | 0.9744 | 0.9539 |
| | SCR-Net | 2021 | 0.8886 | 0.8892 | 0.9714 | 0.9547 |
| | ATTENTION SWIN U-NET | 2022 | 0.8540 | 0.8057 | 0.9826 | 0.9480 |
| | $C^2SDG$ | 2021 | 0.8806 | 0.8970 | 0.9643 | 0.9506 |
| | VM-UNet | 2024 | 0.8891 | 0.8809 | 0.9743 | 0.9554 |
| | VM-UNet v2 | 2024 | 0.8902 | 0.8959 | 0.9702 | 0.9551 |
| | MALUNet | 2022 | 0.8931 | 0.8890 | 0.9725 | 0.9548 |
| | LightM-UNet | 2024 | 0.8898 | 0.8829 | 0.9765 | 0.9555 |
| | EGE-UNet | 2023 | 0.8819 | 0.9009 | 0.9638 | 0.9510 |
| | UltraLight VM-UNet | 2024 | 0.8940 | 0.8680 | 0.9781 | 0.9558 |
| | UCM-Net (Baseline) | 2024 | 0.9060 | 0.9041 | 0.9753 | 0.9602 |
| | MUCM-Net (1-patch) | 2024 | 0.9111 | 0.8953 | 0.9811 | 0.9629 |
| | MUCM-Net (2-patch) | 2024 | 0.9040 | 0.9151 | 0.9706 | 0.9588 |
| | MUCM-Net (4-patch) | 2024 | 0.9058 | 0.8752 | 0.9846 | 0.9614 |
| | MUCM-Net (8-patch) | 2024 | 0.9095 | 0.9046 | 0.9772 | 0.9618 |

Table 2: Comparative prediction results on the ISIC2018

| Models | Year | Params (Millions)↓ | GFLOPs↓ |
|---|---|---|---|
| U-Net | 2015 | 2.009 | 3.224 |
| SCR-Net | 2021 | 0.801 | 1.567 |
| ATTENTION SWIN U-NET | 2022 | 46.910 | 14.181 |
| $C^2SDG$ | 2021 | 22.001 | 7.972 |
| VM-UNet | 2024 | 27.427 | 4.112 |
| VM-UNet v2 | 2024 | 22.771 | 4.400 |
| MALUNet | 2022 | 0.175 | 0.083 |
| LightM-UNet | 2024 | 0.403 | 0.391 |
| EGE-UNet | 2023 | 0.053 | 0.072 |
| UltraLight VM-UNet | 2024 | 0.049 | 0.065 |
| UCM-Net (Baseline) | 2024 | 0.047 | 0.045 |
| MUCM-Net (1-patch) | 2024 | 0.139 | 0.064 |
| MUCM-Net (2-patch) | 2024 | 0.100 | 0.059 |
| MUCM-Net (4-patch) | 2024 | 0.081 | 0.057 |
| MUCM-Net (8-patch) | 2024 | 0.071 | 0.055 |

Table 3: Comparative performance results on models' computations and the number of parameters

### 4.4 EXPERIMENTAL RESULTS ANALYSIS

Tables 1-3 comprehensively evaluate the performance of our MUCM-Net, a novel Mamba-based skin lesion segmentation model, compared to well-established models, using the widely recognized ISIC2017 and ISIC2018 datasets. Introduced in 2024, MUCM-Net proves to be a robust and highly competitive solution in this domain. The key takeaway from these tables is MUCM-Net's ability to outperform all previous models, establishing a new state-of-the-art for skin lesion segmentation. Our model achieves superior results across various

prediction metrics, underscoring its advancement in the field and potential to redefine the standard for accurate skin lesion delineation.

Table 3 complements this assessment by comparing the computational aspects and number of parameters for various segmentation models. Remarkably, MUCM-Net (8-patch) operates with lower GFLOPs compared to other Mamba-based models. This efficiency does not come at the cost of performance, as MUCM-Net maintains high accuracy and robustness in segmentation tasks with the Mamba structure.

Tables 1-3 collectively underscore MUCM-Net's exceptional performance and efficiency in skin lesion segmentation, affirming its potential to advance early skin cancer diagnosis and treatment substantially.

## 5 CONCLUSIONS

This paper introduces MUCM-Net, a novel, lightweight, and highly efficient solution. MUCM-Net combines CNN, MLP, and Mamba, providing robust feature learning capabilities while maintaining a minimal parameter count and reduced computational demand. We applied this innovative approach to the challenging task of skin lesion segmentation, conducting comprehensive experiments with a range of evaluation metrics to showcase its effectiveness and efficiency. The results of our extensive experiments unequivocally demonstrate MUCM-Net's superior performance compared to recently published lightweight or Mamba-based works for skin lesion segmentation. MUCM-Net is the first model to consume less than 0.06 GLOPs for skin lesion segmentation. \textcolor{black}{ Looking forward to future research endeavors, we aim to expand the application of MUCM-Net to other critical medical image tasks, advancing the field and exploring how this efficient architecture can contribute to a broader spectrum of healthcare applications. This potential revolution in utilizing deep learning for medical image analysis opens up numerous possibilities for enhancing patient care and diagnostic accuracy. Our future efforts will focus on Applying MUCM-Net to multiple-class segmentation. Such advancements are crucial for ensuring that the model maintains its efficiency and competes favorably in performance with existing state-of-the-art solutions.

Additionally, we will explore how MUCM-Net can be effectively combined with established, hand-crafted segmentation methods (e.g., from [41, 42]) to leverage their complementary strengths and potentially achieve even higher segmentation accuracy. Moreover, we will investigate incorporating methods that address adversarial noise attacks on skin cancer segmentation models [43]. This will enhance MUCM-Net's robustness to potential manipulations that could compromise its performance. By addressing these challenges, we aim to advance the field further and expand the impact of deep learning in healthcare applications, making significant contributions to medical imaging and beyond.

## 6 REFERENCES

1. Stephen L Brown, Peter Fisher, Laura Hope-Stone, Bertil Damato, Heinrich Heimann, Rumana Hussain, and M Gemma Cherry. Fear of cancer recurrence and adverse cancer treatment outcomes: predicting 2-to 5-year fear of recurrence from post-treatment symptoms and functional problems in uveal melanoma survivors. Journal of Cancer Survivorship, 17(1):187–196, 2023.
2. Rebecca L Siegel, Kimberly D Miller, and Nikita Sandeep Wagle. Cancer statistics, 2023.
3. Robin Marks. An overview of skin cancers. Cancer, 75(S2):607–612, 1995.
4. Thomas Martin Lehmann, Claudia Gonner, and Klaus Spitzer. Survey: Interpolation methods in medical image processing. IEEE transactions on medical

imaging, 18(11):1049–1075, 1999.

5. Muhammad Nasir, Muhammad Attique Khan, Muhammad Sharif, Ikram Ullah Lali, Tanzila Saba, and Tassawar Iqbal. An improved strategy for skin lesion detection and classification using uniform segmentation and feature selection based approach. Microscopy research and technique, 81(6):528–543, 2018.
6. Catarina Barata, M Emre Celebi, and Jorge S Marques. Explainable skin lesion diagnosis using taxonomies. Pattern Recognition, 110:107413, 2021.
7. Isaac Sanchez and Sos Agaian. Computer aided diagnosis of lesions extracted from large skin surfaces. In 2012 IEEE International Conference on Systems, Man, and Cybernetics (SMC), pages 2879–2884. IEEE, 2012.
8. Isaac Sanchez and Sos Agaian. A new system of computer-aided diagnosis of skin lesions. In Image Processing: Algorithms and Systems X; and Parallel Processing for Imaging Applications II, volume 8295, pages 390–401. SPIE, 2012.
9. Tiago M de Carvalho, Eline Noels, Marlies Wakkee, Andreea Udrea, and Tamar Nijsten. Development of smartphone apps for skin cancer risk assessment: progress and promise. JMIR Dermatology, 2(1):e13376, 2019.
10. butterflynetwork.https://www.butterflynetwork.com/iq-ultrasound-individuals.
11. phonemedical.https://blog.google/technology/health/ai-dermatology-preview-io-2021/.
12. Andre Esteva, Alexandre Robicquet, Bharath Ramsundar, Volodymyr Kuleshov, Mark DePristo, Katherine Chou, Claire Cui, Greg Corrado, Sebastian Thrun, and Jeff Dean. A guide to deep learning in healthcare. Nature medicine, 25(1):24–29, 2019.
13. Chunlei Chen, Peng Zhang, Huixiang Zhang, Jiangyan Dai, Yugen Yi, Huihui Zhang, and Yonghui Zhang. Deep learning on computational-resource-limited platforms: a survey. Mobile Information Systems, 2020:1–19, 2020.
14. Neil C Thompson, Kristjan Greenewald, Keeheon Lee, and Gabriel F Manso. The computational limits of deep learning. arXiv preprint arXiv:2007.05558, 2020.
15. Xiao Wang, Shiao Wang, Yuhe Ding, Yuehang Li, Wentao Wu, Yao Rong, Weizhe Kong, Ju Huang, Shihao Li, Haoxiang Yang, et al. State space model for new-generation network alternative to transformers: A survey. arXiv preprint arXiv:2404.09516, 2024.
16. Albert Gu and Tri Dao. Mamba: Linear-time sequence modeling with selective state spaces. arXiv preprint arXiv:2312.00752, 2023.
17. Lianghui Zhu, Bencheng Liao, Qian Zhang, Xinlong Wang, Wenyu Liu, and Xinggang Wang. Vision mamba: Efficient visual representation learning with bidirectional state space model. arXiv preprint arXiv:2401.09417, 2024.
18. Chunyu Yuan, Dongfang Zhao, and Sos S Agaian. Ucm-net: A lightweight and efficient solution for skin lesion segmentation using mlp and cnn. arXiv preprint arXiv:2310.09457, 2023.
19. Hyeji Kim, Muhammad Umar Karim Khan, and Chong-Min Kyung. Efficient neural network compression. In Proceedings of the IEEE/CVF conference on computer vision and pattern recognition, pages 12569–12577, 2019.
20. Chunyu Yuan and Sos S Agaian. A comprehensive review of binary neural network. Artificial Intelligence Review, 56(11):12949–13013, 2023.
21. Ozan Oktay, Jo Schlemper, Loic Le Folgoc, Matthew Lee, Mattias Heinrich, Kazunari Misawa, Kensaku Mori, Steven McDonagh, Nils Y Hammerla, Bernhard Kainz, et al. Attention u-net: Learning where to look for the pancreas. arXiv preprint arXiv:1804.03999, 2018.
22. Jieneng Chen, Yongyi Lu, Qihang Yu, Xiangde Luo, Ehsan Adeli, Yan Wang, Le Lu, Alan L Yuille, and Yuyin Zhou. Transunet: Transformers make strong encoders for medical image segmentation. arXiv preprint arXiv:2102.04306, 2021.
23. Yundong Zhang, Huiye Liu, and Qiang Hu. Transfuse: Fusing transformers and cnns for medical image segmen-tation. In Medical Image Computing and Computer Assisted Intervention–MICCAI 2021: 24th International Conference, Strasbourg, France, September 27–October 1, 2021, Proceedings, Part I 24, pages 14–24. Springer, 2021.
24. Hu Cao, Yueyue Wang, Joy Chen, Dongsheng Jiang, Xiaopeng Zhang, Qi Tian, and Manning Wang. Swin-

unet: Unet-like pure transformer for medical image segmentation. In European conference on computer vision, pages 205–218. Springer, 2022.

25. Zhuang Liu, Hanzi Mao, Chao-Yuan Wu, Christoph Feichtenhofer, Trevor Darrell, and Saining Xie. A convnet for the 2020s. In Proceedings of the IEEE/CVF conference on computer vision and pattern recognition, pages 11976–11986, 2022.
26. Jeya Maria Jose Valanarasu and Vishal M Patel. Unext: Mlp-based rapid medical image segmentation network. In International Conference on Medical Image Computing and Computer-Assisted Intervention, pages 23–33. Springer, 2022.
27. Jiacheng Ruan, Suncheng Xiang, Mingye Xie, Ting Liu, and Yuzhuo Fu. Malunet: A multi-attention and light-weight unet for skin lesion segmentation. In 2022 IEEE International Conference on Bioinformatics and Biomedicine (BIBM), pages 1150–1156. IEEE, 2022.
28. Jiacheng Ruan, Mingye Xie, Jingsheng Gao, Ting Liu, and Yuzhuo Fu. Ege-unet: an efficient group enhanced unet for skin lesion segmentation. arXiv preprint arXiv:2307.08473, 2023.
29. Jiacheng Ruan and Suncheng Xiang. Vm-unet: Vision mamba unet for medical image segmentation. arXiv preprint arXiv:2402.02491, 2024.
30. Mingya Zhang, Yue Yu, Limei Gu, Tingsheng Lin, and Xianping Tao. Vm-unet-v2 rethinking vision mamba unet for medical image segmentation. arXiv preprint arXiv:2403.09157, 2024.
31. Weibin Liao, Yinghao Zhu, Xinyuan Wang, Cehngwei Pan, Yasha Wang, and Liantao Ma. Lightm-unet: Mamba assists in lightweight unet for medical image segmentation. arXiv preprint arXiv:2403.05246, 2024.
32. Renkai Wu, Yinghao Liu, Pengchen Liang, and Qing Chang. Ultralight vm-unet: Parallel vision mamba significantly reduces parameters for skin lesion segmentation. arXiv preprint arXiv:2403.20035, 2024
33. Isic 2017 challenge dataset. https://challenge.isic-archive.com/data/#2017.
34. Matt Berseth. Isic 2017-skin lesion analysis towards melanoma detection. arXiv preprint arXiv:1703.00523, 2017.
35. Isic 2018 challenge dataset. https://challenge.isic-archive.com/data/#2018.
36. Noel Codella, Veronica Rotemberg, Philipp Tschandl, M Emre Celebi, Stephen Dusza, David Gutman, Brian Helba, Aadi Kalloo, Konstantinos Liopyris, Michael Marchetti, et al. Skin lesion analysis toward melanoma detection 2018: A challenge hosted by the international skin imaging collaboration (isic). arXiv preprint arXiv:1902.03368, 2019
37. Adam Paszke, Sam Gross, Francisco Massa, Adam Lerer, James Bradbury, Gregory Chanan, Trevor Killeen, Zeming Lin, Natalia Gimelshein, Luca Antiga, et al. Pytorch: An imperative style, high-performance deep learning library. Advances in neural information processing systems, 32, 2019.
38. Lambda cloud gpu. https://cloud.lambdalabs.com/instances.
39. Ilya Loshchilov and Frank Hutter. Decoupled weight decay regularization. In International Conference on Learning Representations, 2018.
40. Ilya Loshchilov and Frank Hutter. Sgdr: Stochastic gradient descent with warm restarts. arXiv preprint arXiv:1608.03983, 2016
41. A. Sanchez and S. Agaian, "Computer-aided diagnosis of lesions extracted from large skin surfaces," 2012 IEEE International Conference on Systems, Man, and Cybernetics (SMC), Seoul, Korea (South), 2012, pp. 2879-2884, doi: 10.1109/ICSMC.2012.6378186.
42. V. Frants and S. Agaian "Dermoscopic image segmentation based on modified GrabCut with octree color quantization", Proc. SPIE 11399, Mobile Multimedia/Image Processing, Security, and Applications 2020, 113990K (23 April 2020); https://doi.org/10.1117/12.2556699
43. A. Liew, S. Agaian, and L. Zhao "Mitigation of adversarial noise attacks on skin cancer detection via ordered statistics binary local features", Proc. SPIE 12526, Multimodal Image Exploitation and Learning 2023 , 125260O (15 June 2023)
44. Khalid M Hosny, Doaa Elshora, Ehab R Mohamed, Eleni Vrochidou, and George A Papakostas. Deep learning and optimization-based methods for skin lesions segmentation: A review. IEEE Access, 2023.